\def\simless{\mathbin{\lower 3pt\hbox
     {$\rlap{\raise 5pt\hbox{$\char'074$}}\mathchar"7218$}}}   
\def\simmore{\mathbin{\lower 3pt\hbox
     {$\rlap{\raise 5pt\hbox{$\char'076$}}\mathchar"7218$}}}   
\def\msun{{\rm M}_\odot}                                       

\documentclass[useAMS,usenatbib]{mn2e}
\usepackage{epsfig}

\usepackage{ulem}
\title[Spin frequency and kHz QPOs]{Is there a link between the
neutron-star spin and the frequency of the kilohertz quasi-periodic
oscillations?}

\author[M. M\'{e}ndez \& T. Belloni]{Mariano
M\'{e}ndez$^{1,2}$\thanks{E-mail: mariano@sron.nl}, Tomaso
Belloni$^{3}$\\ 
$^{1}$Netherlands Institute for Space Research, Sorbonnelaan 2, 3584 CA
Utrecht, The Netherlands\\
$^{2}$Astronomical Institute Anton Pannekoek, University of Amsterdam,
Kruislaan 403, 1098 SJ Amsterdam, The Netherlands\\
$^{3}$INAF -- Osservatorio Astronomico di Brera,
Via E. Bianchi 46, I-23807 Merate (LC), Italy}

\begin{document}
\normalem

\date{Accepted  Received ; in original form }

\pagerange{\pageref{firstpage}--\pageref{lastpage}} \pubyear{2005}

\maketitle

\label{firstpage}

\begin{abstract}  There is a general consensus that the frequencies of
the kilohertz Quasi-Periodic Oscillations (kHz QPOs) in neutron-star
low-mass X-ray binaries are directly linked to the spin of the neutron
star. The root of this idea is the apparent clustering of the ratio of
the frequency difference of the kHz QPOs and the neutron-star spin
frequency, $\Delta\nu/\nu_s$, at around $0.5$ and $1$ in ten systems for
which these two quantities have been measured. Here we reexamine all
available data of sources for which there exist measurements of two
simultaneous kHz QPOs and spin frequencies, and we advance the
possibility that $\Delta\nu$ and $\nu_s$ are not related to each other.
We discuss ways in which this possibility could be tested with current
and future observations.

\end{abstract}


\section{Introduction}
\label{intro}

Observations of neutron star low-mass X-ray binaries (NS LMXBs) with the
Rossi X-ray Timing Explorer \citep[RXTE;][]{bradt} have led to two
important discoveries: Strong variability on millisecond timescales in
the X-ray light curves of these systems, the so-called kilohertz
quasi-periodic oscillations \citep[kHz QPOs;][]{vanderklis-scox-1-iauc},
and pulsations during X-ray bursts, also known as burst oscillations
\citep{strohmayer-1728-iauc2}.

The kHz QPOs are relatively narrow peaks in the power density spectrum
of NS LMXBs that often appear in pairs, at frequencies $\nu_1$ and
$\nu_2 > \nu_1$ that change with time. These QPOs are thought to reflect
motion of matter at the inner edge of an accretion disk around the
neutron star.

Burst oscillations are short-lived ($\tau \simless 10$s), almost
coherent pulsations seen at the rise and tail of X-ray bursts in NS
LMXBs. The frequency of these oscillations, $\nu_b$, increases in the
tail of the bursts to an asymptotic value that is consistent with being
the same in bursts separated by more than a year time
\citep{strohmayer-longterm}. This, and the fact that in the
accretion-powered millisecond X-ray pulsar (AMP) SAX J1808.4--3658 burst
oscillations appear at the same frequency as the pulsations seen during
persistent (non-burst) intervals \citep{wijnands-1808}, indicates that
the frequency of these burst oscillations is equal to the spin frequency
of the NS, $\nu_s$.

It is commonly accepted that the spin of the neutron star is directly
involved in the mechanism that produces the kHz QPOs. This consensus
stems from the first detection of kHz QPOs and burst oscillations in the
same source, the LMXB 4U 1728--34, very early on in the RXTE mission.
While in different observations the kHz QPOs appeared at different
frequencies, $\nu_1$ in the range $\sim 600 - 800$ Hz, and $\nu_2$ in
the range $\sim 500 - 1100$ Hz, the frequency difference of the QPOs,
when both were present simultaneously, was consistent with being
constant, $\Delta \nu = \nu_2 - \nu_1 \approx 363$ Hz, and also
consistent with the oscillations seen during bursts in  this source at
$\nu_b = 363$ Hz \citep{strohmayer-1728}. This fitted with the
suggestion  \citep{strohmayer-1728-iauc1} that a beat mechanism with the
neutron star spin was responsible for the kHz QPOs. Further results on
other sources \citep[e.g.,][]{ford-0614} appeared to confirm this
picture. A detailed model, the sonic-point model, proposed by
\cite*{miller-1998} explained the observed relation between the kHz QPOs
and the neutron star spin in terms of a beat between material orbiting
at the inner edge of the disk with the Keplerian frequency at that
radius, and the spin of the NS.

As soon as kHz QPOs were discovered in 4U 1636--53
\citep{vanderklis-1636-iauc} with a frequency difference of $\Delta \nu
= 272 \pm 11$ Hz, and burst oscillations at a frequency $\nu_b = 581$ Hz
\citep{zhang-1636-iauc}, it became apparent that in this source $\Delta
\nu$ was inconsistent with being equal to $\nu_b$, but it was close to
$\nu_b/2$. This would have been the end of the sonic-point model, unless
in 4U 1636--53 the 581 Hz frequency seen during X-ray bursts was the
second harmonic of the NS spin frequency, $\nu_b = 2 \times \nu_s$, with
$\nu_s = 290.5$ Hz, e.g. if the pulsed radiation came from two antipodal
poles on the NS. Although searches for a signal at half the burst
oscillations frequency, the putative spin frequency of the neutron star
\citep{miller-sub}, in the power spectrum of the bursts in 4U 1636--53
yielded no positive result \citep{strohmayer-adspr, strohmayer-sub},
this option remained viable.

When kHz QPOs and burst oscillations were discovered in more sources, it
became apparent that there was a systematic trend in how $\Delta \nu$
and $\nu_b$ were related: For sources for which $\nu_b \simless 400$ Hz,
$\Delta \nu \simeq \nu_b$, whereas for sources for which $\nu_b \simmore
400$ Hz, $\Delta \nu \simeq \nu_b/2$. These two groups of sources were
then called ``slow'' and ``fast'' rotators, respectively \citep{muno}.

Related to this, it is interesting to note that when plotted against
each other, the frequencies of the kHz QPOs in 19 different sources all
follow approximately the same relation \citep*{bmh05, bmh07, zhang06}.
This is a priori unexpected if in each source the frequency of the upper
and lower kHz QPOs were related to the spin frequency as $\nu_2 = \nu_1
+ \nu_s$, given that $\nu_1$ and $\nu_2$ span more or less the same
frequency range in all sources of kHz QPOs, whereas the neutron stars in
these systems have spins that span a large range of frequencies, $\nu_s
\approx 200 - 620$ Hz \citep{chakrabarty-1808}.

More kHz QPO data, and more precise QPO frequency measurements, showed
that at least in some sources $\Delta \nu$ was not constant, but
decreased as the QPO frequencies increased \citep{vanderklis-scox-1,
mendez-1608}, and it was always significantly lower than either $\nu_b$
\citep{mendez-1728} or $\nu_b/2$ \citep*{mendez-1636}. Modifications of
the sonic-point model \citep{lamb} could account for this difference,
considering that the frequencies of the QPOs drift slightly when the
material that produces the QPOs crosses the sonic point and falls onto
the neutron star surface.

Three other results raised more serious issues against the sonic-point
beat-frequency model, and eventually rendered it untenable: (i)
\cite*{jonker-1636} found that in 4U 1636--53, when the frequency of the
kHz QPOs decreases sufficiently, $\Delta \nu$ is significantly higher
than $\nu_b / 2$, which was difficult (if not impossible) to explain by
the sonic-point model, even after the modifications introduced by
\cite{lamb}; in the AMP SAX J1808.4--3658, (ii) \cite{chakrabarty-1808}
found that the frequency of burst oscillations is equal to the NS spin
frequency \citep[see][for a similar result in another AMP, XTE
J1814--338]{markwardt-1814-iauc}, while (iii) \cite{wijnands-1808}
detected two simultaneous kHz QPOs with a frequency separation $\Delta
\nu \simeq \nu_s/2$. If this is extended to other sources in which
$\Delta \nu \simeq \nu_b/2$, it would also be true that for those $\nu_b
= \nu_s$, and hence $\Delta \nu \simeq \nu_s/2$. The sonic-point model
could not explain this.

But soon after the SAX J1808.4--3658 results were published, a new model
that reestablished a relation between the spin frequency of the NS and
the kHz QPO, the sonic-point and spin-resonance model
\citep{lamb-miller}, was proposed \citep[see also][]{lee}. In this
model, there is a resonance in the accretion disk at the radial distance
at which the Keplerian orbital frequency is equal to the neutron star
spin frequency minus the vertical epicyclic frequency. This resonance
could lead to either $\Delta \nu = \nu_s$ or $\Delta \nu = \nu_s/2$
depending on whether the disk flow at the resonance radius is smooth or
clumped. In fact, the same source may in principle show both cases, but
this has so far not been observed.

Almost in parallel with some of these explanations, and as a result of
some of the difficulties for beat-frequency models mentioned above, a
different class of models was proposed, in which the frequencies of the
kHz QPOs were associated to two of the three epicyclic frequencies of
general relativity, or a combination of those \citep[e.g.,][]{stella2}.
In these models, the NS spin frequency plays no role in setting up the
frequencies of the kHz QPOs, except for the small corrections it
introduces to the epicyclic frequencies. While these models reproduce
qualitatively the trends seen in the data, and predicted other trends
that were later on observed \citep{migliari-1728, boutloukos-cirx1},
they have problems to fit the data in detail. The main criticism to
these models, however, has always been that they do not explain the fact
that in several sources $\Delta \nu \simeq \nu_s$ or $\Delta \nu \simeq
\nu_s / 2$ \citep{lamb-jvp}. In other words, the criticism is that in
these models the NS spin plays no role in the mechanism that produces
the QPOs. 

Recently, \cite{yin} compared the average frequency separation of the
kHz QPOs, $\langle \Delta \nu \rangle$, with $\nu_s$ in six systems in
which these two quantities have been measured. They suggest that,
despite the low number of sources available for their analysis, $\langle
\Delta\nu \rangle$ depends weakly on $\nu_s$, $\langle \Delta\nu \rangle
\simeq -0.20 \nu_s + 390$ Hz.

In summary, the history of models of the kHz QPOs is a cycle of attempts
to explain the phenomenon in relation to the spin of the NS; each time
that a new observation raised an issue against one such model, a
modification of that model, or a new model, was proposed that tried to
reestablish the role of the NS spin in the mechanism that produces the
kHz QPOs.

After more than ten years from the discovery of the kHz QPOs, a critical
assessment of the current paradigms is necessary. Here we review all the
values of $\Delta\nu$ and $\nu_s$ available in the literature in order
to compare the slow/fast rotator paradigm with other possibilities. We
suggest that the data may in fact show that there is no relation between
NS spin and kHz QPOs. Actually, the data appear to be consistent with a
situation in which the average separation in frequency of the kHz QPOs
$\langle \Delta \nu \rangle$ is more or less constant, independent of
the spin of the neutron star in the system. The division between
``slow'' and ``fast'' rotators may be an effect of the low number of
sources for which two simultaneous kHz QPOs and burst oscillations
and/or pulsations in the persistent emission have been observed, and the
fact that $\langle \Delta \nu \rangle$ is independent of $\nu_s$.

\section{Data}
\label{data}

We use data from the literature. For the rest of the paper we assume
that the frequency $\nu_b$ of burst oscillations is equal to the spin
frequency of the neutron star, $\nu_s$ (see \S\ref{intro}). There are
ten sources for which both $\Delta \nu$ and $\nu_s$ have been measured.
Two of these sources are the AMPs SAX J1808.4--3658 and XTE J1807--294,
six of them are the atoll sources 4U 1608--52, 4U 1636--53, 4U 1702--43,
4U 1728--34, KS 1731--260, and 4U 1915--05 \citep[see][for a definition
of the atoll class]{hk89}, and the other two are IGR J17191--2821 and
SAX J1750.8--2900 which most likely are also atoll sources.

\begin{table}
\caption{Measurements of the frequency separation of the kHz QPOs,
$\Delta \nu$, and spin frequencies, $\nu_s$, for the $2$ AMP and $6+2$
atoll sources studied in this paper. \label{table1}}
\begin{tabular}{lccc}
\hline
Source & $\Delta \nu$ range (Hz) & $\nu_s$ (Hz) & References\\
\hline
\multicolumn{4}{l}{Accretion-powered millisecond X-ray pulsars} \\
\hline
XTE J1807--294    & 180--250 & 191 & 1 \\ 
SAX J1808.4--3658 &   195    & 401 & 2 \\ 
\hline
\multicolumn{4}{l}{Atoll sources}      \\
\hline
4U 1608--52       & 225--325 & 619 & 3 \\ 
4U 1636--53       & 215--330 & 581 & 4 \\ 
4U 1702--43       &   330    & 330 & 5 \\ 
4U 1728--34       & 270--360 & 363 & 6 \\ 
4U 1731--260      &   265    & 524 & 7 \\ 
IGR J17191--2821  &   330    & 294 & 8 \\
SAX J1750.8--2900 &   317    & 600 & 9 \\
4U 1915--05       & 290--355 & 270 & 10\\ 
\hline
\hline
\end{tabular}

REFERENCES ---
(1)  \cite{linares-1807};
(2)  \cite{wijnands-1808};
(3)  \cite{mendez-1608, muno};
(4)  \cite{disalvo-1636, jonker-1636, strohmayer-1636};
(5)  \cite{markwardt-1702};
(6)  \cite{mendez-1728, strohmayer-1728};
(7)  \cite*{wijnands-1731, smith-1731};
(8)  \cite{markwardt-igr, klein-wolt-igr};
(9)  \cite{kaaret-1750};
(10) \cite{boirin-1915, galloway-1915}
\end{table}

For each of these sources we give in Table \ref{table1} the spin
frequency and the range of measurements of $\Delta \nu$. Although there
are too few sources to draw firm conclusions, from this Table it is
apparent that the average of the $\Delta \nu$ range in the AMPs SAX
J1808.4--3568 and XTE J1807--214 is somewhat lower than for the other
eight sources. 

\begin{figure}   
\centerline{\epsfig{file=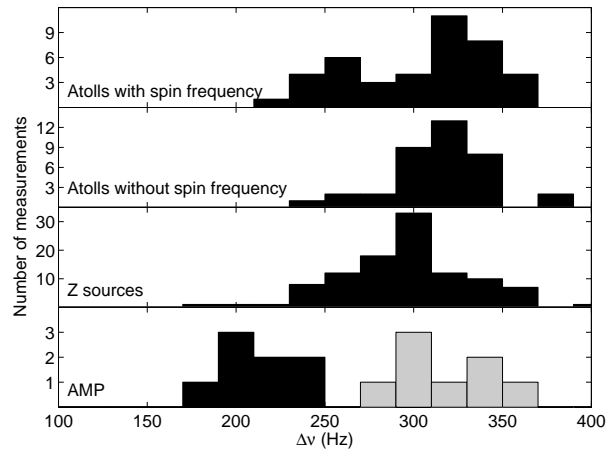,angle=0,width=0.5\textwidth}}  
\caption{The distribution of measurements of $\Delta \nu$ in different
groups of sources. As described in the text, because of the way the
measurements were done, the histograms do not necessarily represent the
true distribution of $\Delta \nu$ (see text for details). {\em Upper
panel:} Atoll sources for which the spin frequency of the neutron star 
is known from burst oscillations. {\em Second panel:} Atoll sources for
which the spin period of the neutron star is not known. {Third panel:} Z
sources; spin frequencies are not known in this case. {\em Lower panel:}
Accreting millisecond X-ray pulsars. The black histogram shows the
actual $\Delta \nu$ measurements; the gray histogram is the distribution
of measurements multiplied by the factors close to 1.5 taken from
\citet{vanstraaten-1808} and \citet{linares-1807}.
\label{distribution}}
\end{figure}

In Figure \ref{distribution} we show in black the distribution of
measurements of $\Delta \nu$ for the ten sources with known spin
frequencies in Table \ref{table1}, and for the other sources of kHz QPOs
for which the spin frequency is not known. The data for this Figure were
taken from the papers in which the kHz QPOs were measured (see van der
Klis 2006 for a complete reference list). We note that to measure both
kHz QPOs significantly and calculate $\Delta \nu$, in most cases power
spectra had to be selected on the basis of some property of the source
\citep[e.g., intensity, colours, characteristic frequency of a low- or
high-frequency timing feature, etc.; see for instance][]{jonker-340+0}
and averaged. In this process, part of the information of the real
distribution of $\Delta \nu$ is lost. Therefore, the plots in Fig.
\ref{distribution} do not show the real $\Delta \nu$ distribution, but
just the range of values observed and a rough idea of how often a
certain value of $\Delta \nu$ has been observed. This plot again
suggests that the average $\Delta \nu$ of the AMP, $\langle \Delta \nu
\rangle \simeq 210$ Hz, is somewhat lower than the average $\Delta \nu$
of the other sources, $\langle \Delta \nu \rangle \simeq 300$ Hz. 

\begin{figure} 
\centerline{\epsfig{file=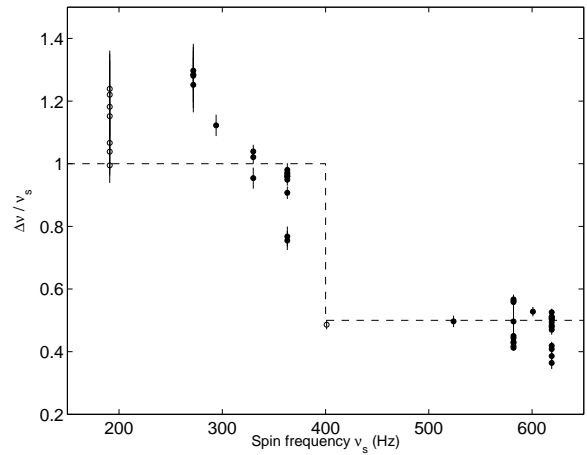,angle=0,width=0.5\textwidth}} 
\caption{The ratio of individual measurements of $\Delta \nu$ divided by
$\nu_s$ as a function of $\nu_s$ for the sources listed in Table
\ref{table1}. The open circles correspond to the two AMPs, and the
filled circles are the other eight sources in Table \ref{table1}. The
dashed line is the step function, $S(\nu_s) = 1$ for $\nu_s \leq 400$
Hz, $S(\nu_s) = 0.5$ for $\nu_s > 400$.
\label{before}}  
\end{figure}

Figure \ref{before} shows the plot of $\Delta \nu / \nu_s$ vs. $\nu_s$
for the ten sources in Table \ref{table1}. For each source we plot all
the individual $\Delta \nu$ measurements, taken from the references
listed in Table \ref{table1}, divided by $\nu_s$ \citep[see ][for a
similar plot]{vanderklis-cefalu}. The open circles correspond to the two
AMPs, and the filled circles are the other eight sources in Table
\ref{table1}. The dashed line in this Figure is a step function,
$S(\nu_s) = 1$ for $\nu_s \leq 400$ Hz, $S(\nu_s) = 0.5$ for $\nu_s >
400$. This Figure shows the fact that for ``slow rotators'', neutron
stars with $\nu_s \simless 400$ Hz, the ratio $\Delta \nu / \nu_s \simeq
1$, whereas for ``fast rotators'', neutron stars with $\nu_s \simmore
400$ Hz, the ratio $\Delta \nu / \nu_s \simeq 0.5$. Note that, as
mentioned in \S\ref{intro}, for some sources $\Delta \nu$ is
significantly different from $\nu_s$ or $\nu_s/2$, respectively, hence
some of the individual ratios are significantly different from 1 or 0.5,
respectively.

\cite*{pbk}, \cite{vanstraaten-0614-1728}, \cite*{vanstraaten-1608},
\cite*{reig-aqlx-1}, \cite*{vanstraaten-1808}, and
\cite{altamirano-1636}, have shown that there is a correlation between
the frequency of the kHz QPOs and that of other low-frequency QPOs. More
specifically, all these authors have shown that when plotted vs. the
frequency of the upper kHz QPO, the frequency of the lower kHz QPO as
well as the frequency of all low-frequency QPOs follow individual
correlations that are consistent with being the same in five atoll
sources, one Z source \cite[see][for the definition of the Z
class]{hk89}, three low-luminosity bursters, and two AMPs
\citep[see][and \citealt{altamirano-1636}, for an overview of these
correlations]{vanstraaten-1808}.

The AMPs SAX J1808.4--3568 and XTE J1807--214 show relations
between the frequencies of the low-frequency QPOs and $\nu_2$ that are
similar to those of the low-luminosity bursters and atoll sources. But
the relations of SAX J1808.4--3568 and XTE J1807--214 are shifted with
respect to those of the other sources \citep{vanstraaten-1808,
linares-1807}. The shift\footnote{This is usually called a shift since
\cite{vanstraaten-1808} and \cite{linares-1807} measure a frequency
shift in a log-log plot.} is between the frequencies of the
low-frequency QPOs and $\nu_2$, and is best described as a
multiplication of $\nu_2$ by a factor close to 1.5. The exact
multiplicative factors are 1.45 for SAX J1808.4--3568 and 1.59 for XTE
J1807--214, respectively. While this factor applied to $\nu_2$ works for
the low-frequency QPOs vs. $\nu_2$ correlations, it does not work for
the correlation between $\nu_1$ and $\nu_2$. Interestingly,
\cite{vanstraaten-1808} and \cite{linares-1807} noticed that the $\nu_1$
vs. $\nu_2$ correlation in the AMPs and in the other sources could be
reconciled if they also multiplied $\nu_1$ by {\em the same factor} that
they used to describe the shift of the $\nu_2$ vs. low-frequency QPO
correlations. (Notice that $\nu_1$ was not used to derive that factor.)

The nature of this shift is unclear, but taken at face value, a
multiplicative factor applied both to $\nu_1$ and $\nu_2$ implies that
the frequency difference $\Delta \nu = \nu_2 - \nu_1$ must also be
multiplied by this factor. The gray histogram in the lower panel of
Figure \ref{distribution} shows the distribution of $\Delta \nu$ (apart
from the caveat described above in this section) for the two AMPs XTE
J1807--294 and SAX J1808.4--3658 multiplied by the factors (close to
1.5) taken from \cite{vanstraaten-1808} and \cite{linares-1807}. This
multiplicative factors appear to bring the values of $\Delta \nu$ in
these two AMP into the range of the values measured in all the other
sources. The average $\Delta \nu$ of the combined sample (Atoll
sources with spin frequency and the two AMPs multiplied by the factors
close to 1.5) is 308 Hz, and the standard deviation is 38 Hz.

\begin{figure} 
\centerline{\epsfig{file=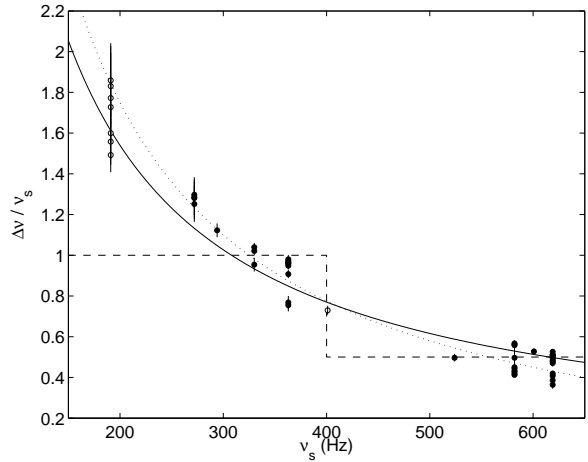,angle=0,width=0.5\textwidth}} 
\caption{The ratio of individual measurements of $\Delta \nu$ divided by
$\nu_s$ as a function of $\nu_s$ for the sources listed in Table
\ref{table1}, but with $\Delta \nu$ of the AMP multiplied by the factors
close to 1.5 from \citet{vanstraaten-1808} and \citet{linares-1807}. The
symbols and the dashed line are same as in Figure \ref{before}. The
solid line corresponds to a constant $\Delta \nu = 308$ Hz. The dotted 
line shows the relation $\Delta\nu = -0.20 \nu_s + 390$ Hz from
\citet{yin}. 
\label{after}} 
\end{figure}

We note that the above procedure could imply a circular argument:
Matching the $\nu_1$ vs. $\nu_2$ correlation of the AMPs and the other
sources via a multiplicative factor means that also $\Delta \nu$ of the
AMPs and of the other sources would match. However, both
\cite{vanstraaten-1808} and \cite{linares-1807} determined the shift
factor on $\nu_2$ using only the correlations between the low-frequency
QPOs and $\nu_2$, independently of $\nu_1$. They then noted that they
could also match the $\nu_1$ vs. $\nu_2$ correlations of the AMPs and
the other sources if they applied the {\em same factor} (within errors)
also to $\nu_1$ \citep[thence, there would be no freedom in choosing the
shift factor on $\nu_1$; see the description in][]{linares-1807}.
Withal, the argument would indeed be circular if the shifts on $\nu_1$
and on $\nu_2$ turned out not to be the same. This conundrum may
eventually be resolved when more shifts in other sources are observed.
In the meantime, we caution the reader about the possible caveats in our
procedure of multiplying $\Delta\nu$ in the AMPs by the factors found by
\cite{vanstraaten-1808} and \cite{linares-1807}.

In Figure \ref{after} we show the plot of  $\Delta \nu / \nu_s$ for the
ten sources in Table \ref{table1}, but now we have multiplied the
$\Delta \nu$ values of the AMPs by the factors taken from
\cite{vanstraaten-1808} and \cite{linares-1807}. As in Figure
\ref{before}, the dashed line is the step function $S(\nu)$ (see above);
the solid line corresponds to a constant $\Delta \nu = 308$ Hz. The
dotted line shows the relation $\Delta\nu = -0.20 \nu_s + 390$ Hz from
\cite{yin}.

\section{Discussion}
\label{discussion}

There is a general (but not universal) tendency to try and include the
spin of the neutron star, $\nu_s$, as a key ingredient in models that
explain the kHz QPOs. In this type of models, the spin is related to the
difference between the frequencies of the kHz QPOs, $\Delta \nu = \nu_2
- \nu_1$. This tendency persisted, despite the fact that several results
seemed to contradict the predictions of these models. Amendments to the
original ideas meant that models had to appeal to rather contrived
geometries to explain the data, and rather artificial classes of sources
had to be introduced to explain the diversity of the results (e.g., the
division of sources with a different relation between $\Delta \nu$ and
$\nu_s$ into ``slow'' and ``fast'' rotators). Here we propose that the
data are in fact consistent with a simpler picture in which the
frequency separation between kHz QPOs, $\Delta \nu$, is independent of
$\nu_s$, with $\langle \Delta \nu \rangle$ more or less constant across
sources.

\begin{figure}   
\centerline{\epsfig{file=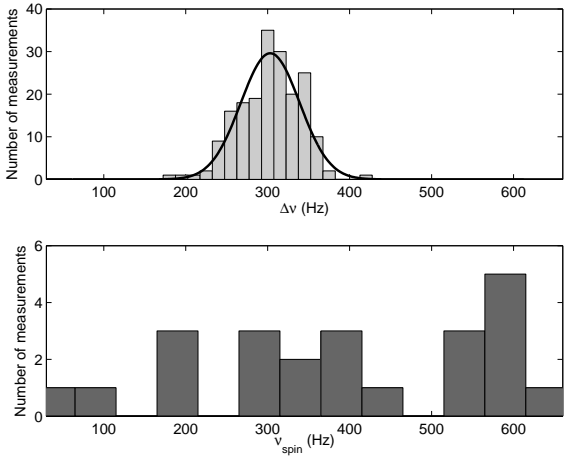,angle=0,width=0.5\textwidth}}
\caption{Upper plot: The distribution of measurements of $\Delta \nu$,
of all types of sources in Fig.\ref{distribution} combined. The solid
line represents the best-fit Gaussian to this distribution, with a
centroid of $303.2 \pm 2.9$ Hz and a standard deviation of $36.0 \pm 2.1$
Hz ($\chi^2 = 14.3$ for 12 degrees of freedom; 1-$\sigma$
errors shown). Lower plot: The distribution of $\nu_s$ in twenty three
sources, nine AMPs with pulsations in the persistent emission and
fourteen sources with burst oscillations \citep[see][and Table
\ref{table1} for the list of sources and spin frequencies]{yin}.
\label{2distributions}}
\end{figure}

In Figure \ref{2distributions} we compare the distribution of the
$\Delta \nu$ measurements in all types of sources for which two
simultaneous kHz QPOs have been detected (see Fig. \ref{distribution}),
with the distribution of spin frequencies of twenty three sources for
which pulsations in the persistent emission or burst oscillations have
been measured \cite[see][and Table \ref{table1} for the list of sources
and spin frequencies]{yin}. From this Figure it is apparent that the
distribution of $\Delta \nu$ measurements is much more concentrated than
the distribution of spin frequencies. The distribution of $\Delta \nu$
measurements can be well described ($\chi^2 = 14.3$ for 12 degrees of
freedom) by a Gaussian with a mean value $\langle \Delta \nu \rangle =
303.2 \pm 2.9$ Hz and a standard deviation $\sigma_{\Delta \nu} = 36.0
\pm 2.1$ Hz (1-$\sigma$ errors). A Kolmogorov-Smirnov (K-S) test yields
a very low probability, $P \approx 7 \times 10^{-7}$, that the two
samples are drawn from the same parent population. According to the
paradigm of slow and fast rotators (see \S\ref{intro}), when $\nu_s
\simmore 400$ Hz (the exact value is not specified) $\Delta \nu$ should
be compared to $\nu_s/2$ instead of $\nu_s$. After dividing by two the
spin frequencies higher than 400 Hz, a K-S test indicates that the
distributions of $\nu_s$  and $\Delta \nu$ are marginally consistent
with each other, with a K-S  probability $P \approx 1 \times 10^{-2}$
that the two are drawn from the same parent population.

Using $\Delta \nu$ and $\nu_s$ for six of the atoll sources also
included in our sample, \cite{yin} proposed that $\Delta \nu$ may be
(weakly) related to $\nu_s$ in a way that is different than predicted by
beat-frequency models. They found that $\langle \Delta\nu \rangle \simeq
-0.20 \nu_s + 390$ Hz\footnote{Notice that to get this result,
\cite{yin} used the individual measurements of $\Delta \nu$ for each
source to calculate $\langle \Delta\nu \rangle$, and the spread of the
$\Delta \nu$ values as a measure of the error. In the case of KS
1731--260 for which there is only one measurement of $\Delta \nu$, they
instead used the error of the measurement of $\Delta \nu$. In the case
of 4U 1702--43 all $\Delta \nu$ measurements were done over a narrow
range of QPO frequencies, which may artificially reduce the spread.}. In
their analysis, \cite{yin} treat the two AMPs differently, and do not
include them in their fit. Following the results of
\cite{vanstraaten-1808} and \cite{linares-1807}, here we do include the
two AMPs in the analysis. We take a step further than \cite{yin}, and we
advance the hypothesis that in fact $\Delta \nu$ and $\nu_s$ are
independent quantities\footnote{Kendall's $\tau$ test \citep[e.g.]
[]{numrec} applied to the data presented here, averaged in the same way
as in \cite{yin}, yields a very low probability (between 1.7 and 1.9
$\sigma$, depending on whether we include the two AMPs are not) that
there is correlation between $\langle \Delta\nu \rangle$ and $\nu_s$}.

Our interpretation that $\Delta \nu$ is independent of $\nu_s$ relies on
the discovery by \cite{vanstraaten-1808} and \cite{linares-1807}, who
found that to reconcile the frequency-frequency correlations of the two
AMPs SAX J1808.4--3568 and XTE J1807--214 with similar correlations in
other sources, the frequencies of the kHz QPOs in the AMPs have to be
multiplied by a factor of $\sim 1.5$ (see \S\ref{data} for a
discussion of possible caveats of this). Both \cite{vanstraaten-1808}
and \cite{linares-1807} find that they can also reconcile the
frequency-frequency correlations if they apply different multiplicative
factors to the frequency of all variability components, except $\nu_2$.
\citep[As noted by][a single multiplicative factor close to 1.5 applied
both to $\nu_1$ and $\nu_2$ without changing the frequency of the other
variability components is the simplest option.]{vanstraaten-1808}. In
particular, they find that if $\nu_2$ remains unchanged, $\nu_1$ of the
AMPs SAX J1808.4--3568 and XTE J1807--214 has to be multiplied by a
factor $\sim 0.8 - 0.9$ to match the $\nu_1 - \nu_2$ correlation defined
by the atoll sources and low-luminosity bursters. The picture presented
here does not change significantly if we only apply the $\sim 0.8 -0.9$
factors to $\nu_1$ and calculate new $\Delta \nu$ values. 

Although it is not the purpose of this paper to explain the nature of
these factors, here we provide some ideas about their possible origin.
The usual suspect is the magnetic field. A stronger field could prevent
the inner edge of the disk from moving inward and, if the kHz QPOs are
produced at that radius, a larger inner disk radius could imply lower
kHz QPO frequencies, $\nu_1$ and $\nu_2$. If, on the contrary, the
low-frequency variability is produced at larger radii, they would be
less affected by the neutron-star magnetic field, which could explain
why a shift of the frequency of the low-frequency components is not
required to match the frequency-frequency correlations. The problem with
this explanation is that other AMPs supposedly having relatively high
magnetic fields, at least comparable to those in SAX J1808.4--3568 and
XTE J1807--214, show no or very small shifts \citep{vanstraaten-1808}.
Also, at least one other non-pulsating source shows shifts in the
correlation, although smaller than the ones in the AMPs:
\cite{altamirano-1820} find a shift of $\sim 1.15$ for the LMXB 4U
1820--30. The other sources showing significant shifts in the
frequency-frequency correlations are the AMPs XTE J0929--314, with
$\nu_s = 185$ Hz \citep{galloway-0929} and a shift of $1.48 \pm 0.11$
\citep{vanstraaten-1808}, and XTE J1814--338 with $\nu_s = 314$ Hz
\citep{markwardt-1814-iauc}, and a shift of $1.21 \pm 0.09$ \citep[][the
shift here is marginally significant]{vanstraaten-1808}.

A neutron-star mass difference could also explain these shifts. E.g., in
the model of \cite{stella2}, and in other models that explain the
frequencies of the kHz QPOs in terms of epicyclic frequencies in general
relativity, the relation between $\nu_1$ and $\nu_2$ depends explicitly
on the neutron-star mass \citep*[see, e.g.,][]{stella3,
boutloukos-cirx1}: As noted by \cite{bmh07}, a multiplicative factor
applied to the neutron-star mass translates into the same multiplicative
factor applied to both kHz QPOs \citep[see eq. (4) in][]{stella3}. If
this (model-dependent) interpretation is correct, a factor $\sim 1.5$ in
$\nu_1$ and $\nu_2$ for the AMPs implies that the neutron stars in those
systems are $\sim 1.5$ times more massive than in the other atoll and Z
sources and the low-luminosity bursters. It is generally assumed that,
due to accretion, the neutron stars in LMXBs have masses larger than the
canonical $1.4 \msun$ neutron star. If the $\sim 1.5$ factor is related
to a difference in neutron-star mass, this would imply uncomfortably
large masses for the neutron stars in the two AMPs. It is somewhat more
difficult to assess the effect of the mass of the neutron star on the
low-frequency components, because it is not clear what frequency in the
model represents the frequency of those components. If one of these
low-frequency QPOs were due to Lense-Thirring precession
\citep{stella1}, its frequency would be \cite[see eq. (1) in][]{stella1}
$\nu_{LT} \propto I M^{-1} \nu^{2}_{2} \nu_s$, where $I$ and $M$ are the
moment of inertia and the mass of the neutron star, respectively, and as
usual $\nu_2$ is the frequency of the upper kHz QPO and $\nu_s$ is the
spin frequency of the neutron star. If this identification is correct,
the spin frequencies in table \ref{table1}, a shift factor $1.5$ in the
mass and in the frequency of the upper kHz QPO, but no shift of the
low-frequency QPOs imply that the moment of inertia must be a factor
$\sim 10$ different among some of these sources.

We know that $\Delta \nu$ is not the same in all sources \citep{zhang06}
and not even for the same source when more than one significant
detection is available \citep{vanderklis-scox-1, mendez-1608}. However,
it is remarkable that, other than in the two AMPs, in all sources in
which two simultaneous kHz QPOs have been detected, $\langle \Delta \nu
\rangle$ is approximately the same. After applying the multiplicative
factors described in \cite{vanstraaten-1808} and \cite{linares-1807},
the same is true for the two AMPs (see \S\ref{data} for possible
caveats). This despite the fact that the measured spin frequencies span
a factor of more than $3$. In the model of \cite{stella2}, $\Delta \nu$
is equal to the radial epicyclic frequency, $\nu_{r}$ which, for the
case of negligible eccentricity and a non-rotating neutron star is
$\nu_{r} = (1-6 GM/r c^2)^{1/2} \nu_{\phi}$, with
$\nu_{\phi}=1/(2\pi)(GM/r^3)^{1/2}$ the azimuthal frequency, identified
with $\nu_2$ in their model. (For neutron stars with spins smaller than
$\sim 600$ Hz and masses in the range $1.4 - 2 \msun$, taking a typical
range of $\nu_2$ frequencies, the radial epicyclic frequency is within
$\approx 15$\% of the value given by this formula.) One would then
expect that on average $\Delta \nu$ would be the same for all neutron
stars if they all have more or less the same mass and their upper kHz
QPO spans more or less the same frequency range. 

If it is generally true that $\langle \Delta \nu \rangle$ is more or
less the same in all sources of kHz QPOs, the idea that there is no link
between $\Delta \nu$ and $\nu_s$ could be tested in the case of the LMXB
EXO 0748--676, which has a spin frequency of $45$ Hz
\citep{villarreal-exo}. From the results in Figure \ref{after}, for EXO
0748--676 one expects $\Delta \nu \approx 300$ Hz, whereas models that
include a direct link between neutron-star spin frequency and
frequencies of the kHz QPOs predict that for this source $\Delta \nu$
should be either $22.5$ Hz or $45$ Hz. In fact, in the context of
``slow'' and ``fast'' rotators, for EXO 0748--676 $\Delta \nu$ is
expected to be $45$ Hz. (We cannot discard that if two simultaneous kHz
QPOs are detected in EXO 0748--676 with $\Delta \nu \approx 300$ Hz,
there would be attempts to modify existing models, or propose completely
new ones, to explain $\Delta\nu / \nu_s$ ratios that are an integer
larger than $1$.) Unfortunately, so far a single kHz QPO has been
observed from this source \citep{homan-exo}. 

An equally interesting test of this idea would be to find a source with
a spin frequency in the range $\nu_s \approx 350 - 500$ Hz for which no
(or only a small) shift is required to fit the frequency-frequency
correlations of \cite{vanstraaten-1808}. A case of more or less constant
$\langle \Delta \nu \rangle$ across different sources implies that
$\Delta\nu/\nu_s$ would be between $0.6$ and $0.8$. Actually, there is a
source that may be used for this in the near future: The AMP XTE
J1751--305 has a spin frequency $\nu_s = 435$ Hz \citep{markwardt-1751},
whereas \cite{vanstraaten-1808} find that a shift of only $1.12\pm 0.03$
applied to $\nu_2$ is required to match the frequency-frequency
correlations. Unfortunately, so far there has been no detection of two
simultaneous kHz QPOs that would allow us to calculate $\Delta\nu$ in
this source.

Despite the fact that the data seem to suggest that $\langle \Delta \nu
\rangle$ is more or less the same in all sources of kHz QPOs (in the
cases of the AMP SAX J1808.4--3658 and XTE J1807--294 after applying a
multiplicative factor deduced from the low-frequency QPO vs. $\nu_2$
correlations; see above and \citealt{vanstraaten-1808} and
\citealt{linares-1807} for details and possible caveats), we have
no strong reason to discard the possibility that there are sources for
which this is not the case (even after applying factors similar to those
deduced in SAX J1808.4--3658 and XTE J1807--294). Our conjecture that
there is no link between $\Delta \nu$ and $\nu_s$ would therefore not be
weakened if a source with two simultaneous kHz QPOs is ever discovered,
for which $\langle \Delta \nu \rangle$ is not close to $\sim 300$ Hz, as
long as in such a source $\langle \Delta \nu \rangle$ (after accounting
for any possible shift factor as the ones in SAX J1808.4--3658 and XTE
J1807--294; \citealt{vanstraaten-1808, linares-1807}) is different from
$\nu_s$ and $\nu_s/2$.

From Figure \ref{after} it is apparent that in SAX J1808.4--3658 and XTE
J1807--294 the shifts on the frequency of the kHz QPOs
\citep{vanstraaten-1808, linares-1807}, and the idea that $\Delta\nu$ is
either equal to $\nu_s$ or $\nu_s/2$ are inconsistent with each other.
It seems unlikely that this issue can be fully resolved as long as the
nature of frequency shifts remains unexplained. While here we propose
that the shifts imply that $\Delta \nu$ is not equal to $\nu_s$ or
$\nu_s/2$, we cannot completely discard that the shifts have a different
explanation, and that in the two AMPs $\Delta\nu / \nu_s$ is indeed
close to either $1$ or $0.5$. We note, however, that even without taking
the shifts into account, there is solid evidence that in several sources
$\Delta\nu$ is significantly different from $\nu_s$ or $\nu_s/2$ (see
\S\ref{intro}). The question is how strong the evidence must be before
the idea that $\Delta\nu$ and $\nu_s$ are directly linked is abandoned.

To conclude, here we put forward the idea that the frequency difference
of the kHz QPOs, $\Delta\nu$, in neutron-star low-mass X-ray binaries
may not be related at all to the spin frequency, $\nu_s$, of the neutron
star. Beat-frequency mechanisms have been proposed not just in the
context of the kHz QPOs; they were originally advanced in the 1980s
\citep{alpar-1985, lamb-1985} to explain the low-frequency QPOs in these
systems. We cannot rule out completely the hypothesis of a similar type
of link between the kHz QPOs and the neutron-star spin, but this idea
can in principle be tested and, if proven wrong, discarded. 

\section*{Acknowledgments}

We thank Diego Altamirano, Jeroen Homan, Peter Jonker, Manuel Linares,
Cole Miller, and Michiel van der Klis for useful comments on earlier
versions of this manuscript, and the referee for his/her excellent
remarks that helped us improve the paper significantly. We also thank
Alice, Amina, Ilaria, and Manuel for their support. TB acknowledges
financial contribution from contract PRIN INAF 2006. The Netherlands
Institute for Space Research (SRON) is supported financially by NWO,
the Netherlands Organisation for Scientific Research. This research has
made use of data obtained through the High Energy Astrophysics Science
Archive Research Center Online Service, provided by the NASA/Goddard
Space Flight Center.

\label{lastpage}


\begin{thebibliography}{99}

\bibitem[\protect\citeauthoryear{Altamirano et al.}{Altamirano et
al.}{2005}]{altamirano-1820} Altamirano, D., van der Klis, M., M\'endez,
M., Migliari, S., Jonker, P. G., Tiengo, A., Zhang, W., 2005, ApJ, 633,
358.

\bibitem[\protect\citeauthoryear{Altamirano et al.}{Altamirano et
al.}{2007}]{altamirano-1636} Altamirano, D., van der Klis, M.,
Klein-Wolt, M., M\'{e}ndez, M., van Straaten, S., Jonker, P. G., Lewin,
W. H. G., Homan, J., 2007, ApJ, submitted

\bibitem[\protect\citeauthoryear{Alplar \& Shaham}{Alpar \&
Shaham}{1985}]{alpar-1985} Alpar, M. A., Shaham, J., 1985, Natur, 316,
249

\bibitem[\protect\citeauthoryear{Belloni, M\'{e}ndez, \& Homan}{Belloni
et al.}{2005}]{bmh05} Belloni T., M\'{e}ndez M., Homan J., 2005, A\&A,
437, 209 

\bibitem[\protect\citeauthoryear{Belloni, M\'{e}ndez, \& Homan}{Belloni
et al.}{2007}]{bmh07} Belloni T., M\'{e}ndez M., Homan J., 2007, MNRAS,
376, 1133 

\bibitem[\protect\citeauthoryear{Bradt, Rothschild, \& Swank}{Bradt et
al.}{1993}]{bradt} Bradt H. V., Rothschild R. E., Swank J. H., 1993,
A\&AS, 97, 355

\bibitem[\protect\citeauthoryear{Boirin et al.}{Boirin et
al.}{2000}]{boirin-1915} Boirin, L., Barret, D., Olive, J. F., Bloser,
P.F., Grindlay, J. E., 2000, A\&A, 361, 121

\bibitem[\protect\citeauthoryear{Boutloukos et al.}{Boutloukos et
al.}{2006}]{boutloukos-cirx1} Boutloukos S., van der Klis M., Altamirano
D., Klein-Wolt M., Wijnands R., Jonker P. G., Fender R. P., 2006, ApJ,
653, 1435 

\bibitem[\protect\citeauthoryear{Chakrabarty et al.}{Chakrabarty et
al.}{2003}]{chakrabarty-1808} Chakrabarty D., Morgan E. H., Muno M. P., 
Galloway D. K., Wijnands R., van der Klis M., Markwardt C. B., 2003,
Natur, 424, 42 

\bibitem[\protect\citeauthoryear{Di Salvo, M\'{e}ndez, \& van der
Klis}{Di Salvo et al.}{2003}]{disalvo-1636} Di Salvo, T., M\'{e}ndez,
M., van der Klis, M., 2003, A\&A, 406, 177 

\bibitem[\protect\citeauthoryear{Ford et al.}{Ford et
al.}{1997}]{ford-0614} Ford, E. C., et al., 1997, ApJ, 475, L123 

\bibitem[\protect\citeauthoryear{Galloway et  al.}{Galloway et
al.}{2002}]{galloway-0929} Galloway D. K., Chakrabarty D., Morgan 
E. H., Remillard R. A., 2002, ApJ, 576, L137 

\bibitem[\protect\citeauthoryear{Galloway et al.}{Galloway et
al.}{2001}]{galloway-1915} Galloway D. K., Chakrabarty D., Muno M. P.,
Savov P., 2001, ApJ, 549, L85 

\bibitem[\protect\citeauthoryear{Hasinger \& van der Klis}{Hasinger \&
van der Klis}{1989}]{hk89}Hasinger, G., \& van der Klis, M., 1989, A\&A,
225, 79

\bibitem[\protect\citeauthoryear{Homan \& van der  Klis}{Homan \& van
der Klis}{2000}]{homan-exo} Homan J., van der Klis M., 2000, ApJ, 539,
847 

\bibitem[\protect\citeauthoryear{Jonker et al.}{Jonker et
al.}{2000}]{jonker-340+0} Jonker, P. G., et al., 2000, ApJ, 537, 374 

\bibitem[\protect\citeauthoryear{Jonker, M\'{e}ndez \& van der
Klis}{Jonker et al.}{2002}]{jonker-1636} Jonker, P. G., M\'{e}ndez, M.,
van der Klis, M., 2002, MNRAS, 336, L1

\bibitem[\protect\citeauthoryear{Kaaret et al.}{Kaaret et
al.}{2002}]{kaaret-1750} Kaaret P., Zand J. J. M., Heise J., Tomsick J.
A., 2002, ApJ, 575, 1018 

\bibitem[\protect\citeauthoryear{Klein-Wolt et  al.}{Klein-Wolt et
al.}{2007}]{klein-wolt-igr} Klein-Wolt M., Wijnands R., Swank
J. H., Markwardt C. B., 2007, ATel, 1075, 1 

\bibitem[\protect\citeauthoryear{Lamb}{Lamb}{2003}]{lamb-jvp} Lamb F.
K., 2003, ASPC, 308, 221 

\bibitem[\protect\citeauthoryear{Lamb et al.}{Lamb et
al.}{1985}]{lamb-1985} Lamb, F. K., Shibazaki, N,. Alpar, M. A., Shaham,
J., 1985, Natur, 317, 681 

\bibitem[\protect\citeauthoryear{Lamb \& Miller}{Lamb \&
Miller}{2001}]{lamb} Lamb, F. K., \& Miller, M. C., 2001, ApJ, 554, 1210

\bibitem[\protect\citeauthoryear{Lamb \& Miller}{Lamb \&
Miller}{2003}]{lamb-miller} Lamb F. K., \& Miller M. C., 2003, ApJ,
submitted, preprint, arXiv:astro-ph/0308179v1 

\bibitem[\protect\citeauthoryear{Lee, Abramowicz, \& Klu{\'z}niak}{Lee
et al.}{2004}]{lee} Lee W. H., Abramowicz M. A., Klu{\'z}niak W., 2004,
ApJ, 603, L93 

\bibitem[\protect\citeauthoryear{Linares et al.}{Linares et
al.}{2005}]{linares-1807} Linares, M., van der Klis, M., Altamirano, D.,
Markwardt, C. B., 2005, ApJ, 634, 1250 

\bibitem[\protect\citeauthoryear{Markwardt et  al.}{Markwardt et
al.}{2007}]{markwardt-igr} Markwardt C. B., Klein-Wolt M., Swank  J. H.,
Wijnands R., 2007, ATel, 1068, 1

\bibitem[\protect\citeauthoryear{Markwardt, Strohmayer \&
Swank}{Markwardt et al.}{1999}]{markwardt-1702} Markwardt, C. B.,
Strohmayer, T., Swank, J. H., 1999, ApJ, 512, L125

\bibitem[\protect\citeauthoryear{Markwardt \&  Swank}{Markwardt \&
Swank}{2003}]{markwardt-1814-iauc} Markwardt C. B., Swank J. H., 2003,
IAUC, 8144, 1  

\bibitem[\protect\citeauthoryear{Markwardt et  al.}{Markwardt et
al.}{2002}]{markwardt-1751} Markwardt C. B., Swank J. H., Strohmayer,
T.E., Zand J. J. M., Marshall F. E., 2002, ApJ, 575, L21 

\bibitem[\protect\citeauthoryear{M\'{e}ndez \& van der Klis}{M\'{e}ndez
\& van der Klis}{1999}]{mendez-1728} M\'{e}ndez, M., \& van der Klis,
M., 1999, ApJ, 517, L51

\bibitem[\protect\citeauthoryear{M\'{e}ndez, van der Klis, \& van 
Paradijs}{M\'endez et al.}{1998}]{mendez-1636} M\'{e}ndez M., van der
Klis M., van Paradijs J., 1998, ApJ, 506, L117 

\bibitem[\protect\citeauthoryear{M\'{e}ndez et al.}{M\'{e}ndez et
al.}{1999}]{mendez-1608} M\'{e}ndez, M., van der Klis, M., Wijnands, R.,
Ford, E. C., van Paradijs, J., 1998, ApJ, 505, L23

\bibitem[\protect\citeauthoryear{Migliari, van der Klis, \& 
Fender}{Migliari et al.}{2003}]{migliari-1728} Migliari S., van der Klis
M., Fender R. P., 2003, MNRAS, 345, L35 

\bibitem[\protect\citeauthoryear{Miller}{Miller}{1999}]{miller-sub}
Miller, M. C., 1999, ApJ, 515, L77

\bibitem[\protect\citeauthoryear{Miller, Lamb, \& Psaltis}{Miller et
al.}{1998}]{miller-1998} Miller, M. C., Lamb, F. K., Psaltis, D. ,1998,
ApJ, 508, 791

\bibitem[\protect\citeauthoryear{Muno et al.}{Muno et al.}{2001}]{muno} 
Muno M. P., Chakrabarty D., Galloway D. K., Savov P., 2001, ApJ, 553,
L157 

\bibitem[\protect\citeauthoryear{Press et al.}{Press et al.}{1992}]
{numrec} Press, W. H., Teukolsky, S. A., Vetterling, W. T., Flannery B.
P., 1992, Numerical Recipes in Fortran: The Art of Scientific Computing,
2nd edn. Cambridge Univ. Press, Cambridge


\bibitem[\protect\citeauthoryear{Psaltis, Belloni, \& van der
Klis}{Psaltis et al.}{1999}]{pbk} Psaltis, D., Belloni, T., van der
Klis, M., 1999, ApJ, 520, 262

\bibitem[\protect\citeauthoryear{Reig, van Straaten \& van der
Klis}{Reig et al.}{2004}]{reig-aqlx-1} Reig, P., van Straaten, S., van
der Klis, M., 2004, ApJ, 602, 918

\bibitem[\protect\citeauthoryear{Smith, Morgan, \& Bradt}{Smith et
al.}{1997}]{smith-1731} Smith D. A., Morgan E. H., Bradt H., 1997, ApJ,
479, L137 

\bibitem[\protect\citeauthoryear{Stella \& Vietri}{Stella \&
Vietri}{1998}]{stella1} Stella, L., \& Vietri, M. 1998, ApJ, 492, L59

\bibitem[\protect\citeauthoryear{Stella \& Vietri}{Stella \&
Vietri}{1999}]{stella2} Stella, L., \& Vietri, M., 1999, Phys. Rev.
Lett, 82, 17

\bibitem[\protect\citeauthoryear{Stella, Vietri, \& Morsink}{Stella et
al.}{1999}]{stella3} Stella L., Vietri M., Morsink S. M., 1999, ApJ,
524, L63 

\bibitem[\protect\citeauthoryear{Strohmayer}{Strohmayer}{2001}]
{strohmayer-adspr} Strohmayer T. E., 2001, AdSpR, 28, 511 

\bibitem[\protect\citeauthoryear{Strohmayer \& Markwardt}{Strohmayer \&
Markwardt}{2002}]{strohmayer-sub} Strohmayer T. E., Markwardt C. B., 
2002, ApJ, 577, 337 

\bibitem[\protect\citeauthoryear{Strohmayer et al.}{Strohmayer et
al.}{1996a}]{strohmayer-1728-iauc2} Strohmayer, T., Zhang, W., Smale,
A., Day, C., Swank, J., Titarchuk, L., Lee, U., 1996a, IAUC, 6387, 2 

\bibitem[\protect\citeauthoryear{Strohmayer, Zhang, \& 
Swank}{Strohmayer et al.}{1996c}]{strohmayer-1728-iauc1} Strohmayer T.,
Zhang, W., Swank, J., 1996c, IAUC, 6320, 1 

\bibitem[\protect\citeauthoryear{Strohmayer et al.}{Strohmayer et
al.}{1998}]{strohmayer-1636} Strohmayer, T. E., Zhang, W., Swank, J. H.,
White, N. E., Lapidus, I., 1998, ApJ, 498, L135

\bibitem[\protect\citeauthoryear{Strohmayer et al.}{Strohmayer et
al.}{1998}]{strohmayer-longterm} Strohmayer T. E., Zhang W., Swank J.
H., Lapidus I., 1998, ApJ, 503, L147 

\bibitem[\protect\citeauthoryear{Strohmayer et al.}{Strohmayer et
al.}{1996b}]{strohmayer-1728} Strohmayer T. E., Zhang W., Swank J. H.,
Smale A., Titarchuk L., Day C., Lee U., 1996b, ApJ, 469, L9 

\bibitem[\protect\citeauthoryear{van der Klis}{van der
Klis}{2005}]{vanderklis-cefalu} van der Klis M., 2005, AIPC, 797, 345 

\bibitem[\protect\citeauthoryear{van der Klis}{van der
Klis}{2006}]{vanderklis-review} van der Klis, M. 2006, in Compact
stellar X-ray sources, ed. W. H. G. Lewin, \& M. van der Klis
(Cambridge: Cambridge Univ. Press), 39

\bibitem[\protect\citeauthoryear{van der Klis et al.}{van der Klis et
al.}{1996b}]{vanderklis-1636-iauc} van der Klis, M., van Paradijs, J.,
Lewin, W. H. G., Lamb, F. K., Vaughan B., Kuulkers E., Augusteijn, T.,
1996b, IAUC, 6428, 2 

\bibitem[\protect\citeauthoryear{van der Klis et al.}{van der Klis et
al.}{1996a}]{vanderklis-scox-1-iauc} van der Klis, M., Swank, J., Zhang,
W., Jahoda, K., Morgan, E., Lewin, W., Vaughan, B., van Paradijs, J.,
1996a, IAUC, 6319, 1 

\bibitem[\protect\citeauthoryear{van der Klis et al.}{van der Klis et
al.}{1997}]{vanderklis-scox-1} van der Klis M., Wijnands, R., Horne, K.,
Chen, W., 1997, ApJ, 481, L97 

\bibitem[\protect\citeauthoryear{van Straaten et al.}{van Straaten et
al.}{2002}]{vanstraaten-0614-1728} van Straaten, S., van der Klis M., Di
Salvo, T., Belloni, T., 2002, ApJ, 568, 912 

\bibitem[\protect\citeauthoryear{van Straaten, van der Klis, \&
M\'{e}ndez}{van Straaten et al.}{2003}]{vanstraaten-1608} van Straaten,
S., van der Klis, M., M\'{e}ndez M., 2003, ApJ, 596, 1155 

\bibitem[\protect\citeauthoryear{van Straaten, Wijnands \& van der
Klis}{van Straaten et al.}{2005}]{vanstraaten-1808} van Straaten, S.,
van der Klis, M., Wijnands, R., 2005, ApJ, 619, 455

\bibitem[\protect\citeauthoryear{Wijnands \& van der Klis}{Wijnands \&
van der Klis}{1997}]{wijnands-1731} Wijnands, R., \& van der Klis, M.,
1997, ApJ, 482, L65

\bibitem[\protect\citeauthoryear{Villarreal \&  Strohmayer}{Villarreal
\& Strohmayer}{2004}]{villarreal-exo} Villarreal A. R., Strohmayer T.
E., 2004, ApJ, 614, L121 

\bibitem[\protect\citeauthoryear{Wijnands et al.}{Wijnands et
al.}{1997}]{wijnands-1636} Wijnands, R., van der Klis, M., van Paradijs,
J., Lewin, W. H. G., Lamb, F. K., Vaughan, B. A., Kuulkers, E., 1997
ApJ, 479, L141

\bibitem[\protect\citeauthoryear{Wijnands et al.}{Wijnands et
al.}{2003}]{wijnands-1808} Wijnands, R., van der Klis, M., Homan, J.,
Chakrabarty, D., Markwardt, C. B., Morgan, E. H., 2003, Natur, 424, 44

\bibitem[\protect\citeauthoryear{Yin et al.}{Yin et al.}{2007}]{yin}
Yin, H. X., Zhang, C. M., Zhao, Y. H., Lei, Y. J., Qu, J. L., Song, L.
M., Zhang, F., 2007, A\&A, in press, arXiv:0705.1431v2 [astro-ph]

\bibitem[\protect\citeauthoryear{Zhang et al.}{Zhang et
al.}{1996}]{zhang-1636-iauc} Zhang, W., Lapidus, I., Swank, J. H.,
White, N. E., Titarchuk, L., 1996, IAUC, 6541, 1 

\bibitem[\protect\citeauthoryear{Zhang et al.}{Zhang et
al.}{2006}]{zhang06} Zhang C. M., Yin H. X., Zhao Y. H., Zhang F., Song
L. M., 2006, MNRAS, 366, 1373 

\end{thebibliography}
\end{document}